\documentclass[intlimits,twoside,a4paper]{article}

\usepackage[cp1251]{inputenc}
\usepackage[eqsecnum]{cmpj3}

\issue{2024}{27}{4}{43702}
\doinumber{10.5488/CMP.27.43702}
\title[Effect of time-dependent sinusoidal electric field on the onset of electroconvection in a viscoelastic fluid layer]%
{Effect of time-dependent sinusoidal electric field on the onset of electroconvection in a viscoelastic fluid layer%
}
\author[C. Rudresha, C. Balaji, V. Vidya Shree, S. Maruthamanikandan]{C. Rudresha\orcid{0000-0002-0958-4220}\refaddr{label1}\thanks{Corresponding author: \email{rudresha\textunderscore maths@sirmvit.edu}.},
        C. Balaji\orcid{0000-0002-3832-935X}\refaddr{label2}, V. Vidya Shree\orcid{0000-0003-1554-8258}\refaddr{label3}, S. Maruthamanikandan\orcid{0000-0001-9811-0117}\refaddr{label4}}
\addresses{
\addr{label1} Sir M Visvesvaraya Institute of Technology, Bangalore-562157, Karnataka, India
\addr{label2} CMR Institute of Technology, Bangalore-560037, Karnataka, India
\addr{label3} SJB Institute of Technology, Bangalore-560060, Karnataka, India
\addr{label4} Presidency University, Bangalore-560077, Karnataka, India
}


\date{Received September 9, 2024, in final form November 21, 2024}

\begin{document}

\maketitle

\begin{abstract}
Time-periodic electric field modulation of a viscoelastic dielectric fluid layer heated from below and cooled from above is examined using an Oldroyd-B type liquid. On the basis of small amplitudes of modulation, the regular perturbation method can be used to calculate the threshold for correction of the critical Rayleigh number. The dielectric constant is assumed to be a linear function of temperature. We show that electric field modulation frequency, electrical, Prandtl number, and viscoelastic parameters are related to the shift in the critical Rayleigh number and the possibility of subcritical convection for low-frequency modulation of the electric field. Rayleigh number, wavenumber, and frequency stability are determined based on free-free isothermal boundary conditions. The dielectrophoretic force is only destabilizing when an electrical field is modulated at a low frequency because it is associated with an unmodulated layer of dielectric fluid. As a result of the stress relaxation parameter in a sinusoidal electric field, the system is destabilized at low frequencies and stabilized at moderate and high frequencies. The effect of strain retardation on mechanical anisotropy is completely opposite. The stability characteristics are illustrated through graphs showing the numerical values of parameters.
%
%
\printkeywords
%
\end{abstract}

\section{Introduction}

Over the past several decades, a lot of researchers have been interested in Newtonian fluid heated from below because it has a variety of industrial uses. The thermal instability of Newtonian fluids under various hydrodynamic and hydromagnetic theories was studied by Chandrasekhar \cite{chandrasekhar2013hydrodynamic}. Due to its multiple applications in climatology, oceanography, drug administration, micro-cooling systems, nanotechnology, etc., a lot of researchers are interested in the study of electrohydrodynamic instability in dielectric fluids. Several researchers utilizing various fluid types have looked at how an ac or dc electric field affects natural convection in a horizontal dielectric fluid layer. Landau’s \cite{landau2} study of the onset of electrohydrodynamic convection focused on a horizontal layer of dielectric fluid. The electrohydrodynamic instability in a viscoelastic liquid layer was studied by Takashima and Ghosh \cite{takashima3}, who discovered that oscillatory modes of instability only occur when the liquid layer thickness is less than about 0.5 mm and that, for such a thin layer, the electrical force is significantly stronger than the buoyancy force. Takashima and Hamabata~\cite{takashima4} looked at the stability of natural convection in the presence of a potential difference in a vertical layer of dielectric fluid. The discussions on the uses of electro-hydrodynamics by Maekawa et al. \cite{maekawa5}, Castellanos~\cite{castellanos6}, and Chen et al. \cite{chen7} were condensed. According to them, electro-thermo-hydrodynamics heat transfer has emerged as a substitute strategy for improving heat transmission.

Rudraiah and Garathri \cite{rudraiah8} studied the effect of modulation on thermal convection in a horizontal fluid layer and Chang et al. \cite{chang9}, on the other hand, focused on the numerous uses of electroconvection and the impact of heat modulation on its onset in a porous medium saturated with a dielectric fluid. The normal mode approach and a truncated approximation of the Fourier series were used by Radhakrishna and Siddheshwar \cite{siddheshwar10} to investigate the linear and weakly nonlinear stability analysis of thermal convection in a dielectric liquid permeated by a vertical, uniform ac electric field. It is discovered that increasing the electric Rayleigh number reduces the amplitudes and, as a result, reduces heat transmission. Nagouda and Maruthamanikandan \cite{nagouda11} investigated the impact of heat radiative transmission on creating electroconvection in a porous material in the presence of a uniform vertical alternating electric field. They made the point that the absorption coefficient of a fluid is constant across all wavelengths and unaffected by the environment. However, several researchers have dealt with the topic in depth, and the expanding body of research in the area is well documented \cite{balaji12, chandrashekar13, shree14, shree15}.

Over the past few years, the research of viscoelastic fluids has gained much significance. This is principally due to the several uses they have in problems such as drilling, food and paper production, and other operations of a similar nature. The boundary layer idea of such fluids is of special relevance since it is used to address a number of technical problems. Green \cite{green16} was the first to delve into the problem of convective instability of a viscoelastic fluid heated from below. Vest and Arpaci \cite{vest17} examined the topic of overstability of a viscoelastic fluid. The study of non-Newtonian fluids has received a lot of interest because of its expanding significance in geophysical fluid dynamics, chemical technologies, and the petroleum industry.

Recently, there has been accomplished a study on pattern generation in viscoelastic fluids as well as the stationary and oscillatory instability thresholds \cite{Li18, li19}. Viscoelastic fluids are substantially less likely to experience the flow instability and turbulence than Newtonian fluids because of a high viscosity of the polymeric fluids. For some time, there has been a typical conviction that oscillatory convection is beyond the realm of possibilities in viscoelastic liquids in sensible exploratory conditions \cite{rosen20, larson21}. The capability producing of viscoelastic convection has been demonstrated by recent investigations on the elastic behavior of individual long DNA strands in buffer solutions. This theory is supported by the discovery of oscillatory convection by Kolodner \cite{kolodner22} in DNA suspensions with an annular shape. Theoretically, these findings revive interest in heat convection in viscoelastic fluids. Othman’s \cite{othman23, othman24} study focused on the fundamentals of stability in a horizontal viscoelastic dielectric liquid layer and a vertical temperature gradient.

Maruthamanikandan \cite{maruthamanikandan25} looked into the issue of electro-convective instability in dielectric fluids when surface tension, viscoelasticity, and internal heat generation were present. Malashetty \cite{malashetty26} looked at the convection stability requirements in an anisotropic binary viscoelastic liquid layer. Mubeen et al. \cite{taj27} used a linear stability analysis to examine the problem of the state of convective instability in a horizontal, nontoxic porous layer saturated with a Maxwell viscoelastic fluid and subjected to a zero-order chemical reaction. Viscoelastic fluid convection and its inherent potential for oscillatory instability are discussed. The fluid in the porous media is shown to be more susceptible to instability when chemical interactions are occurring than when they are absent. In a recent work on the pseudo-one-dimensional electroosmotic flow of a viscoelastic fluid by Abhimanyu et al. \cite{abhimanyu28}, the flow of an Oldroyd-B fluid was comprehensively investigated. Shokri et al. \cite{shokri29} conducted research on the linear stability study of the fingering instability of viscoelastic fluids in a rectangular box. For viscoelastic fluids, the Oldroyd-B model was used. In a planar porous layer saturated by a binary viscoelastic fluid, the impact of viscous dissipation on double-diffusive convective flow was recently examined by Kamalika et al. \cite{roy30}.

Semenov \cite{semenov31} investigated the instability of the liquid dielectric parametric thermal layer under a uniform electric field at free isothermal boundaries. The instability may be proven to exist at electric forces that are critical although frequently shift, and that the orders are much larger than the critical power of a constant electric field. Smorodin et al. \cite{smorodin32} used a harmonically time-varying heat flux normal to the top open surface of a fluid layer to study parametric stimulation of thermoelectric instability. When a liquid dielectric layer is exposed to a transverse temperature gradient and an alternative, controlled electric field, Velarde and Smorodin \cite{velarde33} looked into the convective instability of the layer. Rudresha et al. \cite{rudresha34, rudreshac35, rudresha20236} investigated the impact of electric field modulation on electroconvection in a porous medium. They established that at low frequencies of electric field modulation, the effects of porous medium and electric field modulation are mutually antagonistic.

Rudresha et al. \cite{rudreshart37} studied the time-periodic electric field modulation as well as thermo-electric convection in a dielectric fluid under couple stresses. It is shown that the advent of electroconvection may be accelerated or delayed by adjusting the processes of electric field modulation, electric force, and couple stresses. At the absence of an electric field, no convection occurs because the structure is gravitationally stable. Although buoyancy is stabilizing at this point, when the electric field intensity is strong enough, convection patterns that are very comparable to the well-known {Benard} cells are produced. The effects of viscoelasticity and electric field modulation on electroconvection in an Oldroyd-B model with a dielectric liquid are taken into account in this study.

\section{Mathematical formulation}

\begin{figure}[!b]
	\centerline{\includegraphics{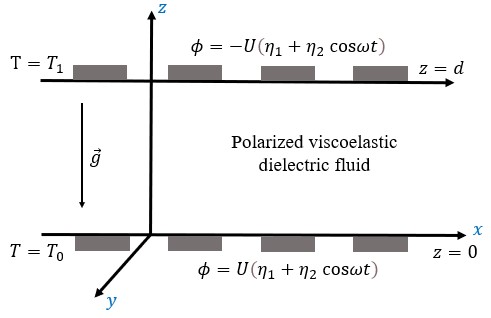}}
	\caption{(Colour online) Schematic diagram.}
	\label{fig-1}
\end{figure}

The investigation described here aims at theoretically investigating the issue of the development of instability in a continuous horizontal layer of viscoelastic dielectric fluid confined between the plates with temperatures ${{T}_{0}}$ and   with modulated electric fields $\phi =\pm U\left( {{\eta }_{1}}+{{\eta }_{2}}\cos \omega t \right)$ (figure~\ref{fig-1}). The respective magnitudes of the constant and reciprocal potential difference components are ${{\eta }_{1}}$ and ${{\eta }_{2}}$, and $U$ is the amplitude of the potential modulation, $\omega $ is the modulation frequency. We simplified the problem by assuming that the dielectric fluid constant is a linear function of temperature, the fluid is incompressible, and the boundary conditions at the walls are free-free, isothermal. This paper examines the liquid prototype designed by Oldroyd \cite{oldroyd38}. Therefore, the constitutive equation is

\begin{align}
	\label{eq1}
      \left( 1+{{\lambda }_{1}}\frac{\partial }{\partial t} \right){{p}_{ij}}=\mu \left( 1+{{\lambda }_{2}}\frac{\partial }{\partial t} \right)\left( \frac{\partial {{v}_{i}}}{\partial {{x}_{j}}}+\frac{\partial {{v}_{j}}}{\partial {{x}_{i}}} \right),
\end{align}
where ${{\lambda }_{1}}$ is the relaxation time, ${{\lambda }_{2}}\,\,\left( <{{\lambda }_{1}} \right)$ is the retardation time, ${{v}_{i}}$ is the velocity, ${{p}_{ij}}$ is the stress deviator, and $\mu $ is the dynamic viscosity. Since we use the theory of linear stability to the fluid layer at rest, the terms of the second order in ${{p}_{ij}}$ and ${{v}_{i}}$ have been neglected in equation (\ref{eq1}). The Boussinesq approximation is used to account for the effects of density variations.

Oldroyd-B elastic-viscous fluid equations for mass, momentum, and thermal energy are \cite{chandrasekhar2013hydrodynamic, rudreshart37, oldroyd38}
\begin{align}
	\label{eq2}
	\nabla \cdot \vec{q}=0,
\end{align}
\begin{align}
	\label{eq3}
	\left( 1+{{\lambda }_{1}}\frac{\partial }{\partial t} \right)\left[ {{\rho }_{0}}\frac{\partial \vec{q}}{\partial t}+\nabla p-\rho \vec{g}+\frac{1}{2}\left( \vec{E}\cdot \vec{E} \right)\nabla \varepsilon  \right]=\left( 1+{{\lambda }_{2}}\frac{\partial }{\partial t} \right)\mu {{\nabla }^{2}}\vec{q},
\end{align}
\begin{align}
	\label{eq4}
	\frac{\partial T}{\partial t}+\left( \vec{q}\cdot \nabla  \right)T=\kappa {{\nabla }^{2}}T,
\end{align}
where $\vec{q}=\left( u,v,w \right),\,\,\vec{g}=\left( 0,0,-g \right),\,\,p,\,\,{{\rho }_{0}},\,\mu ,\,\,T,\,\,\kappa $ and $\vec{E}$ stand for the fluid velocity, gravity acceleration, pressure, density, viscosity, temperature, thermal diffusivity, and the electric field root-mean-square value. The stationary state solution of equations (\ref{eq2}) through (\ref{eq4}) is not time-independent because the dielectrophoretic force is absolutely periodic. We neglected the system's nonlinearities in order to make things straightforward. A charged object in an electric field has a tendency to travel along the lines of the electric field, giving the surrounding fluid velocity. The Maxwell equations are 
\begin{align}
	\label{eq5}
	\nabla \cdot \left[ \varepsilon \vec{E} \right]=0,
\end{align}
\begin{align}
	\label{eq6}
	\nabla \times \vec{E}=0\Rightarrow -\nabla \phi .
\end{align}
The equation of state is
\begin{align}
	\label{eq7}
	\rho ={{\rho }_{0}}\left[ 1-\alpha \left( T-{{T}_{0}} \right) \right],
\end{align}
where $\alpha $ is the thermal expansion coefficient, {$\varepsilon$}  is the dielectric constant, and $\phi$ is the root mean square of the electric potential. It is assumed that the dielectric constant $\varepsilon >0$ is a linear function of temperature and has the following form
\begin{align}
	\label{eq8}
	\varepsilon ={{\varepsilon }_{0}}\left[ 1-e\left( T-{{T}_{0}} \right) \right].
\end{align}

An applied temperature gradient causes electrical conductivity defects in these fluids, and the fluid's electrical conductivity changes with temperature, causing free charges to accumulate in the main body of the fluid. The interaction of these free charges with the applied or produced electric field results in a force that propels the fluid motion.

It is obvious that there exist the following steady solutions:
\begin{align}
	\label{eq9}
	{{T}_{b}}={{T}_{0}}-\beta z,
\end{align}
\begin{align}
	\label{eq10}
	{{\phi }_{b}}=\frac{-2U\left( {{\eta }_{1}}+{{\eta }_{2}}\cos \omega \,t \right)}{\log \left( 1+e\beta d \right)}\log \left( 1+e\beta z \right)+U\left( {{\eta }_{1}}+{{\eta }_{2}}\cos \omega \,t \right),
\end{align}
\begin{align}
	\label{eq11}
	{{E}_{b}}=\frac{2U\left( {{\eta }_{1}}+{{\eta }_{2}}\cos \omega \,t \right)}{d}\left( 1-e\beta z \right).
\end{align}

Let this initial steady state be slightly perturbed so that $\vec{q}=\vec{q}'=\left( u',\,\,v',\,w \right),\,\,p={{p}_{b}}+p',$ $T={{T}_{\,b}}+T',\,\,\varepsilon ={{\varepsilon }_{\,b}}+\varepsilon ',\,\,\phi ={{\phi }_{\,b}}+\phi '$ and $\vec{E}={{\vec{E}}_{b}}+\vec{E}'$. The analysis presented by Chandrasekhar~\cite{chandrasekhar2013hydrodynamic} is closely followed. Substituting these into equations (\ref{eq2}) through (\ref{eq8}), linearizing the equations by neglecting quantities of small order and eliminating the pressure term in the momentum equation, we obtain 
\begin{align}
	\label{eq12}
    \frac{\partial u'}{\partial x}+\frac{\partial v'}{\partial y}=-\frac{\partial w'}{\partial z},
\end{align}
\begin{align}
	\label{eq13}
	{A}_{1} &\left[ 
	{{\rho }_{0}}\frac{\partial }{\partial \,t}\left( {{\nabla }^{2}}w' \right)-\frac{2U{{B}_{\,1}}e{{\varepsilon }_{0}}\beta }{d}\frac{\partial }{\partial z}\left( \nabla _{1}^{2}\phi '  \right) 
	-\alpha {{\rho }_{0}}g\nabla _{1}^{2}T'\right.\nonumber\\
	&\left.-\frac{4{{U}^{2}}B_{\,1}^{\,2}{{e}^{2}}\beta {{\varepsilon }_{0}}}{{{d}^{2}}}\left( \nabla _{1}^{2}T' \right) 
	\right]
	= {A}_{2}\left(\mu {{\nabla }^{4}}w'\right),
\end{align}
\begin{align}
	\label{eq14}
	\frac{\partial T'}{\partial t}-\beta w'=\kappa \,{{\nabla }^{2}}T',
\end{align}
\begin{align}
	\label{eq15}
	{{\nabla }^{2}}\phi '=\frac{-2eU\left( {{\eta }_{\,1}}+{{\eta }_{\,2}}\cos \omega \,t \right)}{d}\frac{\partial T'}{\partial \,z},
\end{align}
where ${A}_{1}=\left( 1+{{\lambda }_{\,1}}\frac{\partial }{\partial t} \right)$,  ${A}_{2}=\left( 1+{{\lambda }_{\,2}}\frac{\partial }{\partial \,t} \right)$,  ${{B}_{1}}={{\eta }_{1}}+{{\eta }_{\,2}}\cos \omega t$ and $\nabla _{1}^{2}=\frac{{{\partial }^{2}}}{\partial \,{{x}^{2}}}+\frac{{{\partial }^{2}}}{\partial \,{{y}^{2}}}$ is the horizontal Laplacian and small terms are neglected using the assumptions $e\beta \ll1$ and $\alpha \beta \ll1$.
Equations (\ref{eq12})--(\ref{eq15}) are first rendered dimensionless by using the scaling $\left( x,\,y,z \right),\,\,T',\,\,t,\,\,w'$ and $\phi '$ by $d,\,\,\Delta T,\,\,{{d}^{2}}/\kappa ,\,\,\kappa /d$ and $2U\left( {{\eta }_{1}}+{{\eta }_{2}}\,f \right)e\Delta T$ as the units of length, temperature, time, velocity, and electric potential, respectively. Using $\Delta T$ as a unit for temperature reduction is intended to simplify the analysis by focusing on relative temperature gradients, which are essential for evaluating the stability conditions in thermal convection. This approach highlights the influence of $\Delta T$ on the critical Rayleigh number, which is central to our study. Neglecting the primes for simplicity, we obtain the linear stability equations in the form

\begin{align}
	\label{eq16}
	\left( 1+\Gamma \frac{\partial }{\partial \,t} \right)\left[ 
		\frac{1}{\Pr }\frac{\partial }{\partial \,t}\left( {{\nabla }^{\,2}}w \right)-{{R}_{e}}B_{\,2}^{\,2}\frac{\partial }{\partial \,z}\left( \nabla _{1}^{2}\phi  \right) 
		-\left[ R+{{\operatorname{R}}_{e}}B_{\,2}^{2} \right]\nabla _{1}^{2}T'
	    \right]=\left( 1+\Gamma \eta \frac{\partial }{\partial \,t} \right){{\nabla }^{4}}w,
\end{align}

\begin{align}
	\label{eq17}
	\left( \frac{\partial }{\partial \,t}-{{\nabla }^{2}} \right)T=w,
\end{align}

\begin{align}
	\label{eq18}
	\left( \nabla _{1}^{2}+\frac{{{\partial }^{2}}}{\partial \,{{z}^{2}}} \right)\phi =-\frac{\partial T}{\partial z},
\end{align}
where  ${{B}_{2}}=1+{{\eta }_{3}}f$, $R=\frac{\alpha \,{{\rho }_{0}}\,g\,{{d}^{4}}\beta }{\mu \,\kappa }$ is the Rayleigh number, ${{R}_{e}}=\frac{4\,{{U}^{2}}{{e}^{2}}{{\beta }^{2}}{{\varepsilon }_{0}}\,{{d}^{2}}\eta _{1}^{2}}{\mu \,\kappa }$  is the electrical Rayleigh number, $\Pr =\frac{\nu }{\kappa }$ is the Prandtl number, ${{\eta }_{3}}=\frac{{{\eta }_{2}}}{{{\eta }_{1}}}$ is the ratio of the amplitude, the parameter $\Gamma =\frac{{{\lambda }_{1}}\,\kappa }{{{d}^{2}}}$ represents the stress relaxation parameter, defined as the ratio of relaxation time to thermal diffusion time and quantifies the fluid’s response to deformation over time. In the Oldroyd-B model, this parameter reflects the extent of stress retention, affecting the fluid's stability under different electric field modulation frequencies, $\eta =\frac{{{\lambda }_{2}}}{{{\lambda }_{1}}}$ is the strain retardation parameter.

Since the fluid layer is contained between two boundaries at the lower surface $z=0$ and the upper surface $z=d$, we now assume that the temperature at the boundaries is kept constant. According to Chandrashekhar \cite{chandrasekhar2013hydrodynamic}, Takashima and Ghosh \citealp{takashima3}, the boundary conditions suitable for the problem are as follows:

\begin{align}
	\label{eq19}
	w={{D}^{\,2}}w=T=\phi =0\quad \text{at}\quad z=0.
\end{align}

Equations (\ref{eq12})--(\ref{eq18}) can be combined to acquire

 \begin{align}
 	\label{eq20}
 	{{L}_{\,3}}\left( \frac{\partial }{\partial t}-{{\nabla }^{\,2}} \right){{\nabla }^{\,4}}w=\left[ R\,{{\nabla }^{\,2}}+{{R}_{\,e}}\nabla _{1}^{\,2}\,{{\left( 1+{{\eta }_{3}}\,f \right)}^{2}} \right]{{L}_{\,1}}\nabla _{1}^{\,2}\,w,
 \end{align}
where  ${{L}_{1}}=\left( 1+\Gamma \frac{\partial }{\partial \,t} \right)$, ${{L}_{2}}=\left( 1+\Gamma \eta \frac{\partial }{\partial \,t} \right)$, ${{L}_{\,3}}={{L}_{\,1}}\frac{1}{\Pr }\frac{\partial }{\partial t}-{{L}_{\,2}}{{\nabla }^{\,2}}$ and $f=\cos \omega t$. 
The boundary conditions can also be expressed in terms of $w$ by making use of equation (\ref{eq17}), which requires $\frac{{{\partial }^{2}}w}{\partial \,{{z}^{2}}}=0$ at the boundaries.

\begin{align}
	\label{eq21}
	w=\frac{{{\partial }^{2}}\,w}{\partial \,{{z}^{2}}}=\frac{{{\partial }^{4}}\,w}{\partial \,{{z}^{4}}}=0\,\,\,\text{at}\,\,z=0,\,\,1 .
\end{align}
We will only look at solutions with a single wave number $\alpha $, such that $\nabla _{1}^{2}w=-{{\alpha }^{2}}\,w$, because the horizontal dependency of $w$ is factorable in this case. Even though the exponential component is omitted for the purpose of clarity in the notation, the use of  ${{\re}^{\ri\left( {{\alpha }_{x}}\,x+{{\alpha }_{y}}\,y \right)}}$ on the horizontal co-ordinates is implicit throughout.

\section{Method of solution}

The Boussinesq equations can be stated in powers of ${{\eta }_{3}}$ due to the small amplitude $\left( {{\eta }_{3}}<1 \right)$ assumption. It implies that the eigenvalues and eigenfunctions of the problem are powers of ${{\eta }_{3}}$ different from those of the traditional Rayleigh number problem of electroconvection instability in a viscoelastic fluid. Therefore, we write

\begin{align} 
\label{eq22}
 	\left( w,R \right)=\left( {{w}_{0}},{{R}_{0}} \right)+{{\eta }_{3}}\left( {{w}_{1}},{{R}_{1}} \right)+\eta _{3}^{2}\left( {{w}_{2}},{{R}_{2}} \right)+\ldots .
\end{align} 

The following system of equations is obtained when the expansions from equation (\ref{eq22}) are substituted into equation (\ref{eq20}) and the powers of ${{\eta }_{3}}$ are partitioned:

\begin{align} 
	\label{eq23}
	L{{w}_{0}},
\end{align} 

\begin{align} 
	\label{eq24}
	L{{w}_{1}}={{R}_{1}}{{L}_{1}}{{\nabla }^{2}}\nabla _{1}^{2}{{w}_{0}}+2{{R}_{e}}{{L}_{1}}f\nabla _{1}^{2}{{w}_{0}},
\end{align} 

\begin{align} 
	\label{eq25}
	L{{w}_{2}}={{R}_{1}}{{L}_{1}}{{\nabla }^{2}}\nabla _{1}^{2}{{w}_{1}}+{{R}_{2}}{{L}_{1}}{{\nabla }^{2}}\nabla _{1}^{2}{{w}_{0}}+2{{R}_{e}}{{L}_{1}}f\nabla _{1}^{2}{{w}_{1}},
\end{align} 
where
\begin{align} 
	\label{eq26}
L={{L}_{3}}\left( \frac{\partial }{\partial t}-{{\nabla }^{2}} \right){{\nabla }^{4}}-{{R}_{0}}{{L}_{1}}\nabla _{1}^{2}{{\nabla }^{2}}-{{R}_{e}}{{L}_{1}}\nabla _{1}^{4}.
\end{align} 

%
Each of ${{w}_{n}}$ is required to satisfy the boundary condition equation (\ref{eq21}). Equation (\ref{eq23}) which is obtained at $O\left( \eta _{3}^{0} \right)$ is the one used in the study of convection in a horizontal dielectric fluid layer or viscoelastic fluid subject to the uniform electric field. The marginally stable solutions for that problem are:
\begin{align} 
	\label{eq27}
w_{0}^{n}=\sin n\piup z.
\end{align} 

On substituting (\ref{eq27}) into (\ref{eq23}), we obtain the following Rayleigh number expression

\begin{align}
	\label{eq28} 
	R_{0}^{n}=\frac{{{\left( {{\alpha }^{2}}+{{n}^{2}}{{\piup}^{2}} \right)}^{4}}-{{R}_{e}}{{\alpha }^{4}}}{{{\alpha }^{2}}\left( {{\alpha }^{2}}+{{n}^{2}}{{\piup }^{2}} \right)}.
\end{align} 

The least eigenvalue for a fixed wave number $\alpha $ occurs when $n = 1$ and it is represented by

\begin{align} 
	\label{eq29}
	{{R}_{0}}=\frac{{{\left( {{\alpha }^{2}}+{{\piup }^{2}} \right)}^{4}}-{{R}_{e}}{{\alpha }^{4}}}{{{\alpha }^{2}}\left( {{\alpha }^{2}}+{{\piup }^{2}} \right)},
\end{align} 
corresponding to ${{w}_{0}}=\sin \piup z$, where ${{\alpha }^{2}}=\alpha _{x}^{2}+\alpha _{y}^{2}$ is the horizontal wave number. The equation for ${{w}_{\,1}}$, then reads
\begin{align}
	\label{eq30} 
	L{{w}_{1}}=2{{R}_{e}}{{\alpha }^{4}}{\Re}\left[ \sum\limits_{n=1}^{\infty }{{{\re}^{\ri\omega t}}\left( 1-\ri\omega \Gamma  \right)\sin \piup z} \right].
\end{align} 

Equation (\ref{eq30}) is homogeneous, and the resonance term makes finding a solution difficult. The right-hand side of this equation must be orthogonal to the operator $L$ null space in order to have a solution for it. This solvability requirement makes it necessary for the time-independent right-hand side to be an orthogonal $\sin \piup z$ in actual situations. ${{R}_{1}}$ is zero since there is just one constant term $-{{R}_{1}}{{\alpha }^{2}}\sin \piup z$ because $f$ fluctuates sinusoidally over time. As a result, all odd coefficients are equal to zero
\begin{align}
	\label{eq31} 
	{{w}_{1}}=2{{R}_{e}}\,{{\alpha }^{4}}{{\operatorname{R}}_{eal}}\left[ \sum\limits_{n=1}^{\infty }{\frac{{{\re}^{-\ri\omega t}}}{L(\omega ,n)}}\left( 1-\ri\omega \Gamma  \right)\sin \piup z \right],
\end{align} 
where $L\left( \omega ,n \right)=-{{C}_{1}}-\ri\omega {{C}_{2}}$ with ${{C}_{1}}$ and ${{C}_{2}}$ are given a bit later. It follows that
\begin{align}
	\label{eq32} 
	L{{\re}^{-\ri\omega t}}\sin n\piup z=L\left( \omega ,n \right){{\re}^{-\ri\omega t}}\sin n\piup z.
\end{align} 
Equation (\ref{eq25}) now reads
\begin{align}
	\label{eq33} 
	L{{w}_{2}}={{R}_{2}}{{L}_{1}}{{\nabla }^{2}}\nabla _{1}^{2}{{w}_{0}}+2{{R}_{e}}{{\alpha }^{4}}{\Re}\left\{ 1-2\ri\omega \Gamma  \right\}f{{w}_{1}}.
\end{align} 
Although the answer to this equation is not necessary, we use it to get ${{R}_{2}}$, the first non-zero modification to $R$. The steady portion of the right-hand side must be orthogonal to $\sin \piup z$ in order to satisfy the solubility requirement. Thus,
\begin{align}
	\label{eq34} 
{{R}_{2}}=\frac{-2{{R}_{e}}{{\alpha }^{2}}}{\left( {{\alpha }^{2}}+{{\piup }^{2}} \right)}\left[ 2{\Re}\left\{ 1-2\ri\omega \Gamma  \right\}\int\limits_{0}^{1}{\overline{f{{w}_{1}}}\sin \piup z\rd z} \right],
\end{align}
where the bar represents the time average. Now, using equation (\ref{eq30}), we obtain
\begin{align}
	\label{eq35} 
	\overline{f{{w}_{1}}}\sin \piup z=\frac{1}{2{{R}_{e}}{{\alpha }^{4}}{\Re}\left\{ 1-\ri\omega \Gamma  \right\}}\overline{{{w}_{1}}L{{w}_{1}}}
\end{align}

The time average is determined by using equations (\ref{eq30}) and (\ref{eq31}) to arrive at $\overline{{{w}_{1}}L{{w}_{1}}}$, and equation~(\ref{eq34}) then produces

\begin{align}
	\label{eq36} 
	{{R}_{2}}=\frac{-2{{R}_{e}}{{\alpha }^{6}}}{\left( {{\alpha }^{2}}+{{\piup }^{2}} \right)}\left[ \sum\limits_{n=1}^{\infty }{\frac{{{C}_{1}}{{C}_{3}}-{{C}_{2}}{{C}_{4}}}{{{\left( {{C}_{1}} \right)}^{2}}+{{\left( \omega {{C}_{2}} \right)}^{2}}}} \right],
\end{align}
where
\begin{align*}
&{{C}_{1}}=\frac{{{\omega }^{2}}}{\Pr }{{\left( {{n}^{2}}{{\piup }^{2}}+{{\alpha }^{2}} \right)}^{2}}+{{\omega }^{2}}\Gamma \left[ \eta +\frac{1}{\Pr } \right]{{\left( {{n}^{2}}{{\piup }^{2}}+{{\alpha }^{2}} \right)}^{2}}-{{\left( {{n}^{2}}{{\piup }^{2}}+{{\alpha }^{2}} \right)}^{4}}+{{R}_{0}}{{\alpha }^{2}}\left( {{n}^{2}}{{\piup }^{2}}+{{\alpha }^{2}} \right)+{{R}_{e}}{{\alpha }^{4}},\nonumber\\
&{{C}_{2}}=\frac{{{\omega }^{2}}\Gamma }{\Pr }{{\left( {{n}^{2}}{{\piup }^{2}}+{{\alpha }^{2}} \right)}^{2}}+{{\left( {{n}^{2}}{{\piup }^{2}}+{{\alpha }^{2}} \right)}^{3}}\left[ 1+\frac{1}{\Pr } \right]-\Gamma \eta {{\left( {{n}^{2}}{{\piup }^{2}}+{{\alpha }^{2}} \right)}^{4}}+{{R}_{0}}\Gamma {{\alpha }^{2}}\left( {{n}^{2}}{{\piup }^{2}}+{{\alpha }^{2}} \right)+{{R}_{e}}\Gamma {{\alpha }^{4}},\nonumber\\
&{{C}_{3}}=\left( 1+2{{\omega }^{2}}{{\Gamma }^{2}} \right),\quad{{C}_{4}}={{\omega }^{2}}\Gamma .
\end{align*}

\section{Results and discussion}

A horizontal viscoelastic dielectric fluid layer is studied to determine how small-amplitude electric field modulation affects hydrodynamic instability. The critical Rayleigh number is determined as a function of the electrical Rayleigh number, Prandtl number, frequency of the modulation, and viscoelastic characteristics using an approximate linear stability analysis proposed by Venezian \cite{venezian39}. The current model assumes a macroscopic layer where continuum mechanics apply. For nanometer-scale layers, molecular effects like surface tension and modified dielectric properties become significant, which are not captured by our model and would require further refinement for accurate predictions. The results of this study are envisaged on the supposition that viscoelasticity dampens the convection currents and that the amplitude of the electric field modulation is minimum enough to allow for the disregard of nonlinear effects. Additionally, the value of the modulating frequency $\omega $ affects the validity of the results. When $\omega $ is small, the electric field modulation affects the entire volume of the fluid, causing significant disruptions. On the other hand, at moderate and high levels of frequency, the consequences of the modulation have a destabilizing impact. This is due to the fact that high frequencies induce the static electric field to renormalize, which causes the electric force to assume a mean value and give rise to the unmodulated equilibrium state.

\begin{figure}[htb]
	\centerline{\includegraphics[scale=0.4]{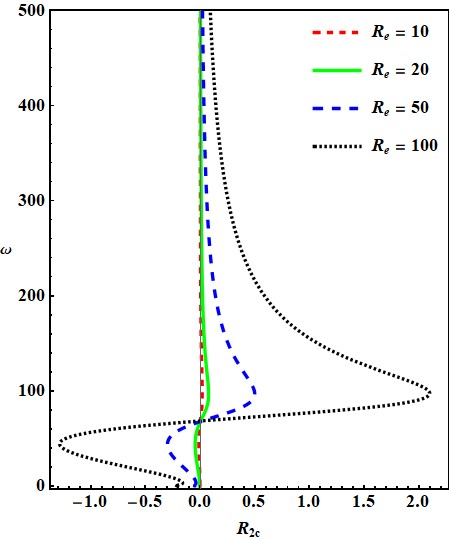}}
	\caption{(Colour online) Plot of ${{R}_{2c}}\,\text{vs.}\,\omega $ for different values of ${{R}_{e}}$ , when $\Gamma =0.7$, ${{\eta }}=0.5$ and $\Pr =50$.}
	\label{fig-2}
\end{figure}

The critical value of the thermal Rayleigh number ${{R}_{2c}}$ and wavenumber $\alpha $ is calculated using the regular perturbation approach, which is based on small amplitudes of modulation. The frequency modulation, thermal Rayleigh number, electrical Rayleigh number ${{R}_{e}}$, Prandtl number $\Pr $, and viscoelastic characteristics ${{\eta }_{v}}$ and $\Gamma $ are used to compute the expression of the critical correction Rayleigh number, and its impact on the stability of the system is examined. The thermal Rayleigh number $R$ effectively describes the magnitude of the buoyancy force caused by the temperature gradient. Physically, the Prandtl number $\Pr $ is the ratio of the viscous and thermal diffusivities, and the electric Rayleigh number ${{R}_{e}}$ is the balance of the energy released by electric force in relation to energy dissipation through viscous friction and thermal dissipation. The stress relaxation parameter $\Gamma $ calculates how much less stress is created in a material when an equivalent amount of strain is produced. The distortion of a material component parts as a result of the applied stress is known as strain retardation ${{\eta }_{v}}$. Both an external force and internal alterations have the potential to cause this deformation. The critical correction Rayleigh number ${{R}_{2c}}$ describes the modulation effects that are both stabilizing and destabilizing. Compared to a system with no modulation, a positive  ${{R}_{2c}}$ shows that the modulation effect is stabilizing the system, whereas a negative ${{R}_{2c}}$ suggests that the modulation impact is destabilizing the system. The findings from the modulated electric field of an electroconvection on a layer of viscoelastic dielectric fluid are presented below.

\begin{figure}[htb]
	\centerline{\includegraphics[scale=0.4]{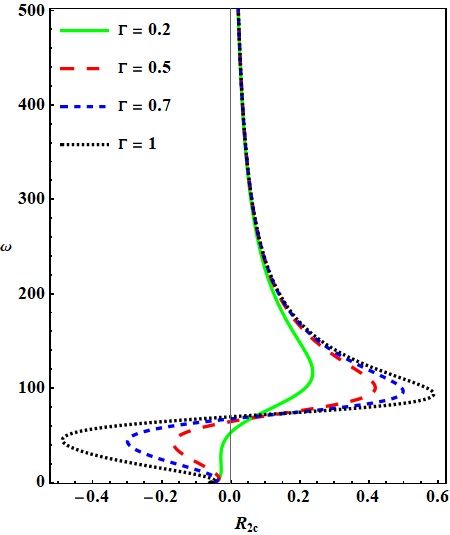}}
	\caption{(Colour online) Plot of ${{R}_{2c}}\,\text{vs}\,\omega $ for different values of $\Gamma $ , when ${{R}_{e}}=50$, ${{\eta }_{v}}=0.5$ and $\Pr =50$.}
	\label{fig-3}
\end{figure}

\begin{figure}[htb]
	\centerline{\includegraphics[scale=0.4]{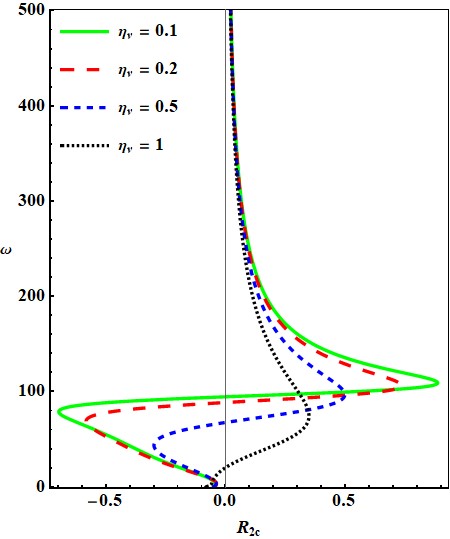}}
	\caption{(Colour online) Plot of ${{R}_{2c}}\,\text{vs}\,\omega $ for different values of ${{\eta }_{v}}$ , when $\Gamma =0.7$, ${{R}_{e}}=50$ and $\Pr =50$.}
	\label{fig-4}
\end{figure}

\begin{figure}[htb]
	\centerline{\includegraphics[scale=0.4]{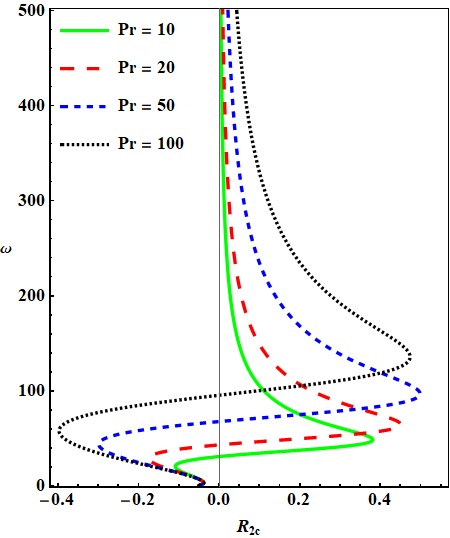}}
	\caption{(Colour online) Plot of ${{R}_{2c}}\,\text{vs}\,\omega $ for different values of $\Pr $ , when $\Gamma =0.7$, $\eta =0.5$ and ${{R}_{e}}=50$.}
	\label{fig-5}
\end{figure}

In figures \ref{fig-2} through \ref{fig-5}, the variation of critical correction Rayleigh number ${{R}_{2c}}$ with frequency $\omega $ for different governing parameters is depicted for the case of electric field modulation. From these figures we find that at the small frequencies, the critical correction Rayleigh number ${{R}_{2c}}$ is negative, indicating that the effect of electric field modulation is destabilizing. On the other hand, for moderate and high frequencies, the critical correction Rayleigh number ${{R}_{2c}}$ is positive indicating that the effect of electric field modulation is stabilizing the system. Figure~\ref{fig-2} shows the variation of ${{R}_{2c}}$ with $\omega $ for a different value of the electrical Rayleigh number ${{R}_{e}}$. This figure shows that for small values of frequency, ${{R}_{2c}}$ becomes negative indicating that the electric field modulation has a destabilizing effect. For moderate and high frequencies, ${{R}_{2c}}$ becomes positive indicating that the electric field modulation diminishes the critical value of Rayleigh number for the onset of electroconvection compared to unmodulated case. It is also observed from figure~\ref{fig-2}, that the electrical Rayleigh number ${{R}_{e}}$ decreases indicating that the effect of ${{R}_{e}}$ is to decline the onset of convection for a small value of frequency. However, for moderated and large values of frequency, the reverse trend is followed for the onset of electroconvection.

Figure~\ref{fig-3} displays the effect of stress relaxation parameter $\Gamma $ on the correction Rayleigh number ${{R}_{2c}}$ for fixed values of electrical Rayleigh number, strain retardation parameter and Prandtl number. It is observed from figure~\ref{fig-3} that the magnitude of ${{R}_{2c}}$ increases negatively with increasing $\Gamma $, indicating that the stress relaxation parameter has destabilizing effect of electric field modulation in a system consisting of a viscoelastic fluid layer for small values of frequency, and we notice that $\Gamma $ has a dual impact in that it exhibits a strong destabilizing effect for small values of $\omega $, whereas for moderate values of $\omega $, it displays a strong stabilizing effect. Furthermore, it should be mentioned that the range of frequency over which the electric field modulation influences the stability criterion increases with the stress relaxation parameter.

The effect of strain retardation parameter ${{\eta }_{v}}$ for the electric field modulation on the stability of the system is shown in figure~\ref{fig-4}. It is seen that as  ${{\eta }_{v}}$  increases, the value of ${{R}_{2c}}$ decreases negatively, indicating that the effect of ${{\eta }_{v}}$ reduces the destabilizing effect of the electric field modulation for small values of the frequency modulation. Moreover, we observe the less destabilizing nature of viscoelastic dielectric fluid compared to the viscous dielectric fluid layer~\cite{rudresha34}. When ${{\eta }_{v}}=0.1$, it has a more stabilizing effect for large values of frequency. Furthermore, the trend gets reversed (which means destabilizing) for small values of the frequency.

Figure~\ref{fig-5} displays variations of ${{R}_{2c}}$ through $\omega $ for a changed value of $\Pr $. Analogous to the influence of viscoelastic parameters ${{\eta }_{v}}$ and $\Gamma $, it is perceived that $\Pr $ also shows a dual impact. Small frequency values of $\omega $ lead to a significant destabilizing effect. We observe a strong stabilization effect for moderate and large frequency values of $\omega $. The effect of $\Pr $ on the electric field modulation diminishes the onset of electroconvection for large values of the frequencies.

\section{Conclusion}

The following results are reached after studying how an electric field modulation affects the onset of electroconvection in a horizontal viscoelastic dielectric fluid layer:

\begin{enumerate} 

\item[1]  When the electric field is modulated at low frequencies, subcritical convective motion is conceivable, but when the electric field is modulated at intermediate or high frequencies, only supercritical convective motion is possible.
\item[2] The classical destabilizing effect of the dielectrophoretic force associated with the unmodulated dielectric fluid layer is only realized for low frequency modulation of the electric field.
\item[3] The effect of the stress relaxation parameter on a time-varying sinusoidal electric field is to destabilize the system at low frequencies and to stabilize it at moderate and high frequencies. Compared to mechanical anisotropy, the strain retardation parameter has the exact opposite effect.
\item[4] The strain retardation parameter diminishes the destabilizing result of modulation in the situation of small values of frequency. However, it reduces stabilization in the case of large values of frequencies.
\item[5] The Prandtl number measures the relationship between viscous force and thermal force; when the Prandtl number rises, the system's electric field modulation exhibits a magnifying behavior regardless of the frequency range.
\item[6] All of the governing parameters play a significant role in the effective control of electroconvection subjected to modulation of a viscoelastic dielectric fluid layer by means of a time-varying electric field. 
\end{enumerate}
The study demonstrates that in the presence of viscoelastic Oldroyd-B fluid, the electric field modulation may induce or delay electroconvection in a viscoelastic dielectric fluid layer. Therefore, the electric field modulation may be used to regulate convective the instability in a horizontal layer of viscoelastic dielectric fluids.


\bibliographystyle{cmpj}
\bibliography{cmpjxampl}

\ukrainianpart

\title{Вплив залежного від часу синусоїдального електричного поля на виникнення електроконвекції у шарі в'язкопружної рідини}
\author[С. Рудреша, С. Баладжі, В. Від'я Шрі, С. Марусаманікандан]{С. Рудреша\refaddr{label1},
	С.~Баладжі\refaddr{label2}, В.~Від'я Шрі\refaddr{label3}, С.~Марусаманікандан\refaddr{label4}}
\addresses{
	\addr{label1} Технологічний інститут ім. М. Вісвесварая, Бангалор-562157, Карнатака, Індія
	\addr{label2} Технологічний інститут CMR, Бангалор-560037, Карнатака, Індія
	\addr{label3} Технологічний інститут SJB, Бангалор-560060, Карнатака, Індія
	\addr{label4} Президентський університет, Бангалор-560077, Карнатака, Індія
}

\makeukrtitle

\begin{abstract}
\tolerance=3000%
Періодичну в часі модуляцію електричного поля шару в’язкопружного діелектричного плину, нагрітого знизу та охолодженого зверху, досліджено за допомогою моделі рідини типу Oldroyd-B. Для малих амплітуд модуляції, метод регулярних збурень може застосовуватися для розрахунку порогу корекції критичного числа Релея. Вважається, що діелектрична проникність є лінійною функцією температури. Показано, що частота модуляції електричного поля, електричні та в’язкопружні параметри, а також число Прандтля пов’язані зі зсувом критичного числа Релея та можливістю субкритичної конвекції для низькочастотної модуляції поля. Число Релея, хвильове число та стійкість частоти визначаються на основі вибраних ізотермічних граничних умов. Діелектрофорезна сила має дестабілізуючу природу лише тоді, коли електричне поле модулюється на низькій частоті, оскільки воно пов’язане з немодульованим шаром діелектричної рідини. Внаслідок релаксації параметра напружень в синусоїдальному електричному полі, система дестабілізується при низьких частотах і стабілізується при помірних та високих частотах. Зауважимо, що вплив уповільнення деформації на механічну анізотропію є цілком протилежним. Характеристики стійкості системи проілюстровані рисунками, що відображають числові значення параметрів.
\keywords конвекція, діелектрична рідина, електричне поле, модуляція, модель Oldroyd-B
\end{abstract}
\end{document}